\documentclass[aps,prl,superscriptaddress,showpacs,twocolumn,citeautoscript]{revtex4}

\usepackage{graphicx}
\usepackage{amsmath}
\usepackage{amssymb}

\newcommand{\bea}{\begin{eqnarray*}}
\newcommand{\eea}{\end{eqnarray*}}
\newcommand{\bne}{\begin{equation*}}
\newcommand{\ede}{\end{equation*}}

\newcommand{\bnen}{\begin{equation}}
\newcommand{\eden}{\end{equation}}
\newcommand{\bean}{\begin{eqnarray}}
\newcommand{\eean}{\end{eqnarray}}
\newcommand{\bsen}{\begin{subequations}}
\newcommand{\esen}{\end{subequations}}

\newcommand{\bna}{\begin{array}}
\newcommand{\eda}{\end{array}}
\newcommand{\bnm}{\begin{enumerate}}
\newcommand{\edm}{\end{enumerate}}

\renewcommand{\vec}[1]{\text{\boldmath{$ #1 $}}}

\newcommand{\avg}[1]{\langle #1 \rangle}

\newcommand{\Dso}{\Delta_{\rm SO}}

\begin{document}
\title{Disorder-mediated electron valley resonance in carbon nanotube quantum dots}

\author{Andr\'as P\'alyi}
\affiliation{Department of Physics, University of Konstanz, D-78457 Konstanz, Germany}
\affiliation{Department of Materials Physics, E\"otv\"os University Budapest, 
H-1517 Budapest POB 32, Hungary}

\author{Guido Burkard}
\affiliation{Department of Physics, University of Konstanz, D-78457 Konstanz, Germany}

\date{\today}

\begin{abstract}
We propose a scheme for coherent rotation of the valley isospin of a single
electron confined in a carbon nanotube quantum dot. 
The scheme exploits the ubiquitous atomic disorder of the nanotube crystal lattice, which induces time-dependent
valley mixing as the confined electron is pushed back and forth along the nanotube 
axis by an applied ac electric field.
Using experimentally determined values for the disorder strength we estimate 
that valley Rabi oscillations with a period on the nanosecond
timescale are feasible.
The valley resonance effect can be detected in the electric current through a double quantum dot in the single-electron transport regime. 
\end{abstract}

\maketitle


\emph{Introduction.} 
The conduction and valence bands of graphene and carbon nanotubes (CNTs) 
form two valleys as the bands approach each other at two non-equivalent points
(K and K') of the Brillouin zone \cite{Saito-cntbook}.
This twofold degeneracy of the electronic spectrum implies that 
the valley degree of freedom of an electron
 can be regarded as one bit of classical or quantum
information, in analogy with the electron spin \cite{Rycerz-valleytronics}.
To evaluate the potential of encoding information in the valley degree of freedom,
it is necessary to explore the physical mechanisms which could provide control 
over the valley state, as well as those leading to the loss of information.

Recent theoretical proposals suggest the use of valley-polarized
edge states of zigzag graphene nanoribbons in a valley filter 
device \cite{Rycerz-valleytronics}, and non-adiabatic magnetic field sweeps for
manipulation of the valley state of a single electron in a graphene Aharonov-Bohm
ring \cite{Recher-graphenering}. 
However, the uncontrolled and strong valley mixing due to 
edge irregularities of nanostructured graphene
might pose a significant challenge towards the realization of these ideas.
This difficulty motivates the study of valley physics in CNTs, which have
a rolled-up and therefore edge-free geometry.
In fact, high-quality CNT quantum dots (QDs) have been fabricated
recently \cite{Kuemmeth-spinorbit-in-cnts,Churchill-13cntprl,Steele-cntdqd}
 and the fourfold (spin and valley) quasi-degeneracy of the QD energy levels
as well as very weak disorder have been experimentally 
confirmed \cite{Kuemmeth-spinorbit-in-cnts,Churchill-13cntprl}.
These ultraclean QD devices offer a chance for experimental realization of 
spin control via spin-orbit interaction \cite{Bulaev-socincntdots} and
manipulation of a combined spin-valley qubit (the ``Kramers qubit'') utilizing
a magnetic field, valley mixing, and bends in a 
CNT \cite{Flensberg-bentnanotubes}.

\begin{figure}
\includegraphics[scale=0.34]{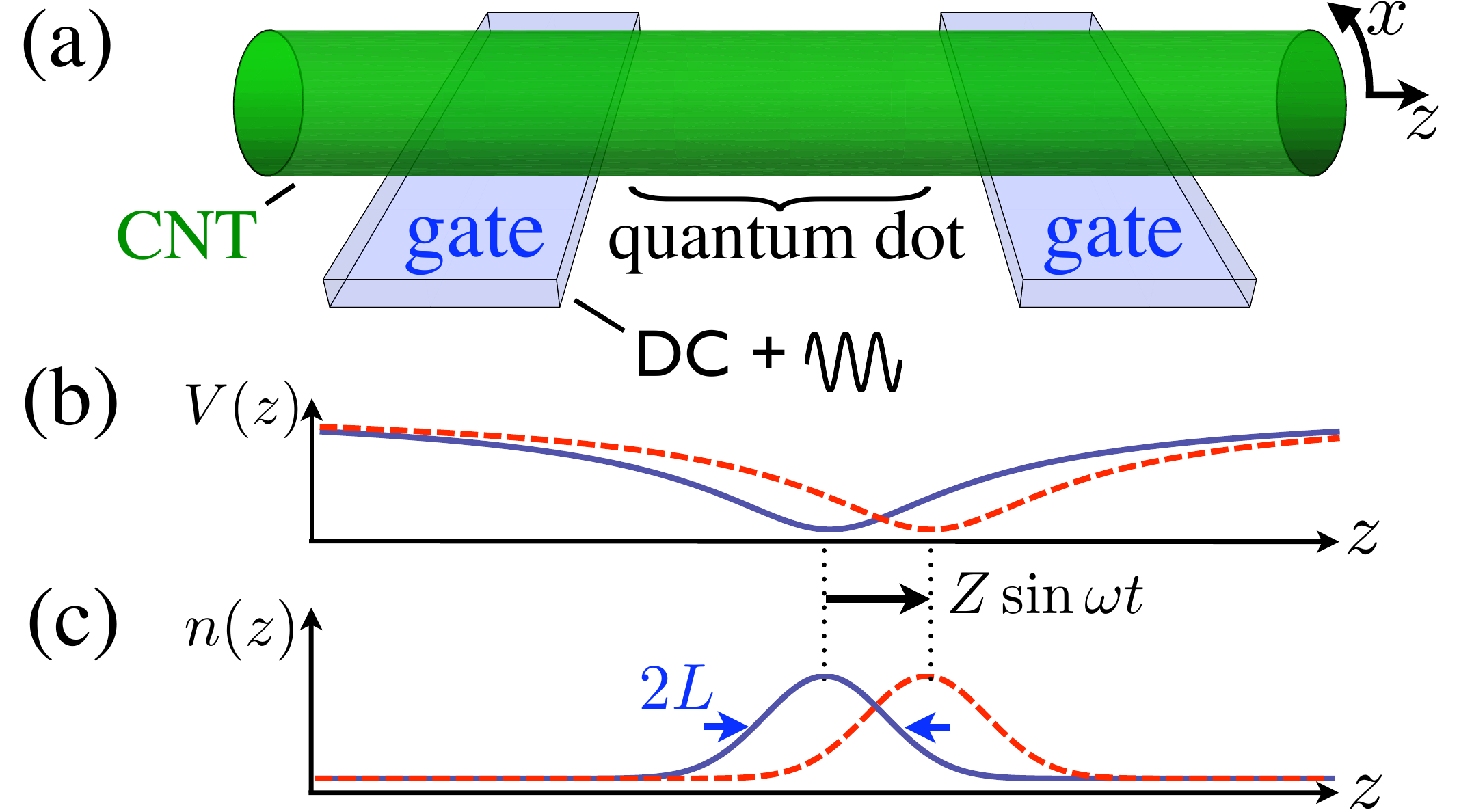}
\caption{\label{fig:singledot}
(Color online)
Schematic of the valley resonance setup. 
(a) Two gate electrodes close to the CNT confine an electron
to the QD region. 
(b) The confinement potential $V(z)$ is periodically modulated in time 
by an ac voltage on the left gate.
(c) The modulation of the confinement potential is followed by 
a displacement of the
electron density $n(z)$.
The interplay of this driven motion and the disorder on the CNT induces
ac valley mixing leading to the valley resonance effect.}
\end{figure}

In this work we propose a scheme for coherent control of the valley isospin of
a single electron,
which is confined in a QD in a straight CNT at zero magnetic field.
The effect, which we name \emph{electron valley resonance} (EVR)
in analogy to electron spin resonance (ESR),
relies on short-range atomic  
disorder (substitutionals, adatoms, vacancies)
of the CNT crystal lattice.
Disorder induces harmonic time-dependent
valley mixing as the confined electron is pushed back and forth along the CNT
by an applied oscillating (ac) electric field (Fig. \ref{fig:singledot}),
which in turn induces Rabi oscillations between the two valley states.
The ac electric field can be induced by applying an ac
voltage component
on one of the confining electrodes, as demonstrated in
single-spin control experiments in GaAs QDs \cite{Nowack-esr,Laird-prl-edsr}.

Using a microscopic model \cite{ando-aob1,Palyi-cnt-spinblockade}
of disorder-induced valley mixing,
we map the valley dynamics of the oscillating electron to
the two-level ESR problem.
Together with a statistical treatment of possible disorder configurations,
we find that the ratio between
typical energy scales of the ac and static valley mixing is $Z/L$, 
where $Z$ is the displacement amplitude of the electron and $2L$ is the width of
the electronic wave function  [see Eq.~\eqref{eq:bac}]. 
As the static valley-mixing energy is 
measurable \cite{Kuemmeth-spinorbit-in-cnts,Churchill-13cntprl},
our finding can be used to estimate the ac component and hence
the frequency of the valley Rabi oscillations.
For a numerical example (see below) using data from 
\cite{Kuemmeth-spinorbit-in-cnts}, with $Z=2$ nm and $L=50$ nm
we find that the time needed for a half Rabi oscillation between the
two valley states is $\approx 1.6$ ns.
We also describe a setup that could be used to detect the EVR
effect via a measurement of the electric current in the regime of 
single-electron transport through a CNT double QD (DQD).

\emph{Microscopic theory of EVR.}
To model the electronic states we use the envelope function 
approximation and take the coordinates defined in Fig.~\ref{fig:singledot}.
The Hamiltonian includes four terms: 
(i) the kinetic energy \cite{ando-aob1}
$H_{\rm kin} = v_{\rm F} (\tau_3 \sigma_1 p_x + \sigma_2 p_z)$
with Fermi velocity $v_{\rm F}$, circumferential (longitudinal) electron
momentum $p_x$ ($p_z$), and $\sigma_i$ ($\tau_i$) being Pauli matrices
acting in sublattice (valley) space;
(ii) curvature-enhanced spin-orbit interaction \cite{Ando-spinorbit,Kuemmeth-spinorbit-in-cnts,Jeong-soi-cnt,Izumida-soiincnt} 
$H_{\rm so} = \tau_3 s_z (\Delta_0 + \sigma_1 \Delta_1)$
with axial spin component $s_z$ and chirality-dependent
on-site (off-site) spin-orbit matrix element
$\Delta_0$ ($\Delta_1$);
(iii) QD confinement potential $V_0(z)$,
and 
(iv) disorder induced by atomic defects \cite{ando-aob1} which randomize the
on-site energies on the crystal lattice,
\bnen
H_{\rm dis}(\vec r) = 
\Omega_{\rm cell} \sum_{l} \sum_{\sigma = A,B} \sigma_\sigma
 [\tau_0 + \tau_r(\varphi_{l\sigma})]
U_{l\sigma} \delta(\vec r - \vec r_{l\sigma}),
\eden
where $U_{l\sigma}$ is the on-site energy on site $\sigma \in (A,B)\equiv (+,-)$
in unit cell $l$.
Here
$\Omega_{\rm cell}$ is the area of the unit cell of the graphene lattice,
$\sigma_{A,B} = (\sigma_0 \pm \sigma_3)/2$,
$\tau_r(\varphi) = \cos \varphi \tau_1 + \sin\varphi \tau_2$,
$\varphi_{l\sigma} = 
\sigma \left( 2\vec K \cdot \vec r_{l\sigma} - \eta \right) +\delta_{\sigma,B} 2\pi/3$, 
$\eta$ is the chiral angle of the CNT,
$\vec r_{l\sigma}$ is the position of the lattice site $l\sigma$,
and $\vec K$ is the vector pointing to the K point of the Brillouin zone.
The presence of the off-diagonal valley operator $\tau_{r}$ in $H_{\rm dis}$
reflects the fact that atomic disorder, due to its short-range character, allows for large
momentum transfer upon scattering, including intervalley
(K $\leftrightarrow$ K') transitions.
Our valley rotation scheme relies on this K-K' coupling. 

EVR is induced by pushing the confined electron back and forth 
along the $z$ axis by applying an ac voltage on one of the gates 
(Fig.~\ref{fig:singledot}). 
We model this time-dependent displacement with the confinement
potential $V(z,t) = V_0(z-Z\sin\omega t)$
with the ac frequency $\omega/2\pi$.
To map the system to a two-level ESR problem, we first transform
the complete Hamiltonian into the reference frame co-moving
with the displaced electron. 
Thus we perform the time-dependent unitary transformation 
$U(t) = e^{iZ\sin(\omega t) p_z/\hbar}$, which leaves the position-independent
kinetic energy and spin-orbit Hamiltonians $H_{\rm kin}$ and $H_{\rm so}$
invariant, 
but renders the confinement potential time-independent 
$V(z,t) \mapsto V_0(z)$ and the disorder term time-dependent
$H_{\rm dis} (\vec r) \mapsto H_{\rm dis}(x,z+Z\sin\omega t)$.
The additional 
term $i\hbar \dot U(t) U^{-1}(t)$, arising from the time-dependence of 
the transformation, can be neglected 
as it does not induce transitions within the ground state QD level.

As the next step, we restrict our consideration to the  
fourfold-degenerate (spin and valley) QD ground-state.
For our purposes we can estimate the corresponding 
eigenfunctions of $H_{\rm kin}+V_0(z)$
as Gaussians, having equal weight on the two sublattices:
$
\Psi_{K,s}(x,z) \equiv {\mid \!\! K \rangle} \chi_s = \frac{e^{iqx}}{\sqrt{2\pi R}} (G_L(z),G_L(z),0,0) \chi_s$, and
$\Psi_{K',s}(x,z) \equiv {\mid \!\! K' \rangle} \chi_s = \frac{e^{-iqx}}{\sqrt{2\pi R}} (0,0,G_L(z),G_L(z))\chi_s$,
with spinors $\chi_s$ having 
spin projection $s \in (\uparrow,\downarrow) \equiv (+,-)$ along
the $z$ axis, circumferential wave number $q>0$,
CNT radius $R$, and
$G_L(z) = \frac{ e^{-z^2/2L^2} } {\pi^{1/4}\sqrt{2L}} $.
Projecting the spin-orbit Hamiltonian to this four-fold 
degenerate subspace yields
$\bar H_{\rm so} = \frac 1 2 \Dso s_z \left(
\mid \!\! K' \rangle \langle K' \!\! \mid - 
\mid \!\! K \rangle \langle K \!\! \mid \right)$,
where $\Dso=2(\Delta_0 + \Delta_1)$;
experimentally reported 
values \cite{Kuemmeth-spinorbit-in-cnts,Churchill-13cntprl,Jhang-cntspinorbit} 
of $\Dso$ are in the range 0.17-2.5 meV.
The form of $\bar H_{\rm so}$ implies that spin-orbit interaction induces an energy
splitting $\Dso$ between 
two valley states having the same spin direction.
Therefore, as in the case of ESR in CNT QDs \cite{Bulaev-socincntdots},
no magnetic field is needed for EVR .
Henceforth we restrict our considerations to the limit of small displacements
$Z/L \ll 1$.
By projecting the disorder Hamiltonian to the four-dimensional 
subspace of interest we find
\bnen
\label{eq:hdiseff}
\bar H_{\rm dis}  = 
\left(b e^{i\phi} + b_{\rm ac} e^{i\phi_{\rm ac}} \sin \omega t \right) \!\!
\mid \!\! K' \rangle \langle K \!\! \mid \!\!  + {\rm h.c.},
\eden
where valley-diagonal terms are omitted.
The real quantites $b$, $\phi$, $b_{\rm ac}$ and $\phi_{\rm ac}$ describe
disorder-induced static and ac valley mixing and can be expressed
in terms of $U_{l\sigma}$ and $\Psi_{v,s}$.
Randomness of the disorder configuration $U_{l\sigma}$
implies the randomness of those quantities as well.
Assuming a homogeneous and uncorrelated 
distribution of the atomic defects 
$\avg{U_{l\sigma} U_{l'\sigma'}}=\avg{U_{l\sigma}^2}\delta_{l\sigma,l'\sigma'}$,
with zero average $\avg{U_{l\sigma}} =0$, 
we find $\avg{b} = \avg{b_{\rm ac}} = 0$ and
\bean
\avg{b^2}& =&
\frac 1 {4 \sqrt 2 \pi^{3/2}} \frac{\Omega_{\rm cell}}{RL} \avg{U_{l\sigma}^2},\\
\label{eq:bac}
\avg{b_{\rm ac}^2}& =& \left(\frac Z L \right)^2 \avg{b^2}.
\eean

Equations \eqref{eq:hdiseff} and \eqref{eq:bac} are the central results of this work.
The ac valley-mixing term $\propto b_{\rm ac}$  in Eq. \eqref{eq:hdiseff},
induced by the simultaneous presence of the ac electric field and atomic disorder,
allows for coherent rotations of the valley isospin similarly to a transverse
magnetic field in ESR
\endnote{Note that here the ac valley-mixing field is not strictly transversal
to the static Hamiltonian if $b$ is finite. However, if $b \ll \Dso$, as found
in recent experiments \cite{Kuemmeth-spinorbit-in-cnts,Churchill-13cntprl},
then the ac field is transversal to a good approximation.}.
Furthermore, using the remarkably simple expression in Eq. \eqref{eq:bac} we can 
estimate the corresponding Rabi frequency from a measurement of the static
valley-mixing matrix element $b$. 
For example, in \cite{Kuemmeth-spinorbit-in-cnts} 
a spin-orbit splitting of $\Dso = 370\, \mu$eV and a valley gap of 
$\Delta_{KK'}=65\, \mu$eV was 
found. 
Identifying $\Delta_{KK'}$ with $2 \sqrt{\avg{b^2}}$,
and taking $L=50$ nm and $Z=2$ nm, from Eq.~\eqref{eq:bac} 
we find an estimate for the strength of the ac 
valley-mixing term in this sample as
$\sqrt{\avg{b_{\rm ac}^2}} = 1.3\, \mu$eV.
This value translates to a $\pi \hbar / \sqrt{\avg{b^2_{\rm ac}}} \approx 1.6\, $ns 
long half Rabi cycle between two orthogonal valley states at resonant driving
$\hbar \omega = \sqrt{\Dso^2+4b^2}$.

\begin{figure}
\includegraphics[scale=0.18]{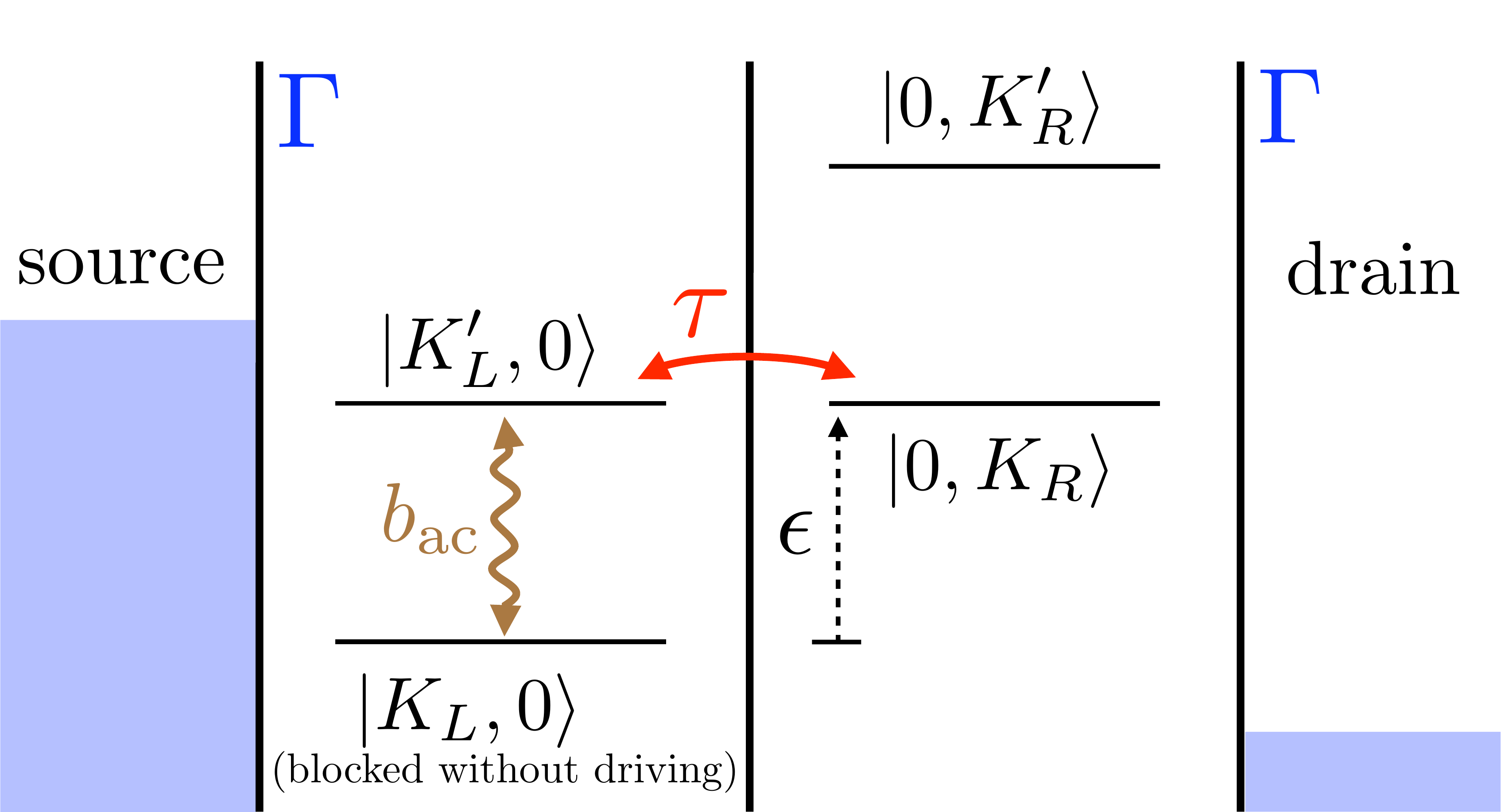}
\caption{\label{fig:doubledot}
(Color online)
Schematic of EVR detection.
The disorder-mediated ac valley mixing
 drives transitions between the states on the left QD.
Interdot tunneling between $K'_L$ and $K_R$ has the
amplitude $\tau e^{i\phi_\tau}=t \langle  K_R \!\! \mid \!\! K'_L \rangle$.
}
\end{figure}

\emph{Detection.}  We now describe a setup where EVR
could be detected (Fig. \ref{fig:doubledot}).
A serially coupled DQD between a source and a drain lead
is tuned appropriately (see below for details), such that 
during the single-electron transport process the electron
can be trapped in a specific valley state in the left QD. 
($\mid \!\! K_L,0 \rangle$ in Fig.~\ref{fig:doubledot}).
The valley state of a trapped electron is changed due to the EVR mechanism
when an ac electric field is applied to the left QD.
This allows the particle to exit via the ground state level 
of the right QD, thus EVR is detected via the measurement
of the electric current through the DQD.
Since this method is based on single-electron transport, it is
unaffected by strong correlations, unlike 
the Pauli blockade effect 
\cite{Churchill-13cntprl,Churchill-cntspinblockade,Palyi-spinblockade,Palyi-cnt-spinblockade}.

To describe the transport process outlined above, 
from now on we consider only spin-$\uparrow$ electrons; hence the spin-orbit
Hamiltonian simplifies to $\bar H_{\rm so} = \frac 1 2 \Dso \left(
\mid \!\! K' \rangle \langle K' \!\! \mid - 
\mid \!\! K \rangle \langle K \!\! \mid \right)$.
The description of the spin-$\downarrow$ electrons is completely analogous.
As shown in Fig. \ref{fig:doubledot}, the energy levels in the DQD are
tuned by gate electrodes so that the higher-lying sublevel 
in the left QD is aligned with the lower-lying sublevel in the
right QD.
We consider single-electron transport via the $(0,0)\to (1,0) \to (0,1) \to (0,0)$
transport cycle where $(n,m)$ refers to the charge configuration
with $n$ ($m$) electrons in the left (right) QD.
Interdot tunneling is assumed to be spin- and valley-conserving, therefore
the static Hamiltonian of a single electron in the DQD is
\bnen
H_{\rm DQD} = \left(\bna{cccc}
   -\frac \Dso 2 & b_L e^{-i\phi_L} & t & 0 \\
   b_L e^{i \phi_L} & \frac \Dso 2 & 0 & t \\
   t & 0 & -\frac \Dso 2 + \epsilon & b_R e^{-i \phi_R} \\
   0 & t & b_R e^{i \phi_R} & \frac \Dso 2 +\epsilon   
 \eda\right),
\eden
where we use the basis $\mid \!\! K,0\rangle$, $\mid \!\! K',0 \rangle$,
$\mid \!\! 0,K\rangle$, $\mid \!\! 0,K' \rangle$, and $L,R = 0,K,K'$
in $\mid \!\! L,R \rangle$ refer to the occupation and valley state in the left and right
QDs.
The left-right energy detuning, needed for the the level alignment described
above, is $\epsilon = (\sqrt{\Dso^2+4b_L^2}+
\sqrt{\Dso^2+4b_R^2})/2$.
For the sake of simplicity and clarity, from now on we assume 
$b_{L,R} \ll \Dso$ (as found in recent
experiments \cite{Kuemmeth-spinorbit-in-cnts,Churchill-13cntprl}),
and $t \ll \Dso$ (tunability of the interdot tunneling has been demonstrated in 
CNT DQDs \cite{Mason-cntdqds}).
A crucial point for our detection scheme is
that the electron interacts with a different set of impurities in the
two different QDs, and therefore $b_L$ and $\phi_L$ are in general different from 
$b_R$ and $\phi_R$.
Together with the condition $b_{L,R} \ll \Dso$,
this implies that the higher-lying energy eigenstate  in 
the left QD $\mid \!\! K'_L \rangle$ has a small $K$ component,
the lower-lying state in the right QD $\mid \!\! K_R \rangle$ 
has a small $K'$ component, and $\mid \!\! K'_L \rangle$ is typically 
not orthogonal to $\mid \!\! K_R \rangle$.
This leads to a finite tunneling matrix element 
$\tau e^{i\phi_\tau} = t \langle K_R \!\! \mid \!\! K'_L \rangle$
between the two aligned levels $\mid \!\! K'_L,0 \rangle$ and
$\mid \!\! 0,K_R \rangle$.
This nonzero matrix element is required for  the EVR detection
scheme we outline below.  

The level structure shown in Fig.~\ref{fig:doubledot} results in 
a blockade effect in a finite source-drain voltage bias.
If the incoming electron enters the QD in the higher-lying $\mid \!\! K'_L,0 \rangle$
state then it can move through the DQD easily because of the complete
hybridization of $\mid \!\! K'_L,0 \rangle$ with $\mid \!\! 0,K_R \rangle$.
However, the condition $t \ll \Dso$ ensures that hybridization
of the lower-lying state $\mid \!\! K_L,0 \rangle$ with (0,1) states is small 
($\lesssim t/\Dso$ in amplitude);
therefore an electron occupying that state 
blocks transport as it exits the DQD slowly. 
Assuming equal tunneling rates $\Gamma$ at source and drain, 
from a transport model based on the secular approximation 
($\Gamma \ll t,b_{L,R},\Dso$) of the 
Born-Markov master equation we estimate that the current through 
the DQD without ac driving 
is much smaller than barrier transparency $\Gamma$,
\bnen
\label{eq:i0}
I_0/ e \approx 2 \Gamma (t/\Dso)^2 \ll \Gamma.
\eden

This transport blockade allows for detection of the EVR
effect described above.
If an ac electric field is active in the left QD and pushes back and forth the electron within the left dot,
occupying the blocking lower-lying state
 $\mid \!\! K_L,0\rangle$ in Fig.~\ref{fig:doubledot},
then it undergoes a Rabi transition to the higher-lying  $\mid \!\! K'_L,0\rangle$ state,
which allows it to exit the DQD to the drain due to the strong hybridization 
of $\mid \!\! K'_L,0 \rangle$ and $\mid \!\! 0,K_R \rangle$.
This process is most effective around resonant driving, i.e., when 
$ 
\hbar \omega \approx  \sqrt{\Dso^2+4b_L^2}.
$
In fact, below we show that around the resonance condition and an appropriate
tuning of the tunneling amplitude $\tau$ the current through the DQD approaches
its maximal value $2e\Gamma/7$. 

To provide an analytical result for the ac-field-induced current we neglect
perturbative hybridization amplitudes 
between (1,0) and (0,1) states which are $\lesssim t/\Dso$.
In this simplified picture the higher-lying level $\mid \!\! 0,K'_R\rangle$ 
on the right QD is not involved in the transport problem as it is not
hybridized with (1,0) states.
After a transformation into the rotating frame in the left QD and the counter-rotating
frame in the right QD, and within the rotating wave approximation (RWA),
the dynamics in the remaining three-level system can be
described by
\bean
\label{eq:hrot}
H_{\rm RWA} &\approx& \delta \mid \!\! K_L,0 \rangle \langle K_L,0\!\! \mid \\
  &+& \left(
  \frac{b_{\rm ac}}{2} 
  \mid \!\! K'_L,0 \rangle \langle K_L,0 \!\! \mid +
  \tau \!\! \mid \!\! 0, K_R \rangle \langle K'_L,0 \!\! \mid
  +{\rm h.c.}\right) \nonumber
\eean
with the ac valley-mixing term $\propto b_{\rm ac}$ 
and the detuning parameter
$\delta = \hbar \omega - \sqrt{\Dso^2+4b_L^2}$.
The phases $\phi_{\rm ac}$ and $\phi_\tau$ have 
been eliminated by a specific choice of the basis states.
A small term $\propto b_{\rm ac} b /\Dso$ has been neglected in
Eq.~\eqref{eq:hrot}.
The leading-order (in $\Gamma$) analytical result for the current, 
obtained from the Born-Markov master equation
transport model is
\bnen
\label{eq:resonantcurrent}
\frac{I}{e\Gamma} =
 \frac{2 \tau^2 b_{\rm ac}^2}
 {4 \tau^4+\tau^2(3 b_{\rm ac}^2 -8 \delta^2)+4 \delta^4+
 5 \delta^2 b_{\rm ac}^2+b_{\rm ac}^4}.
\eden

\begin{figure}
\includegraphics[scale=0.22]{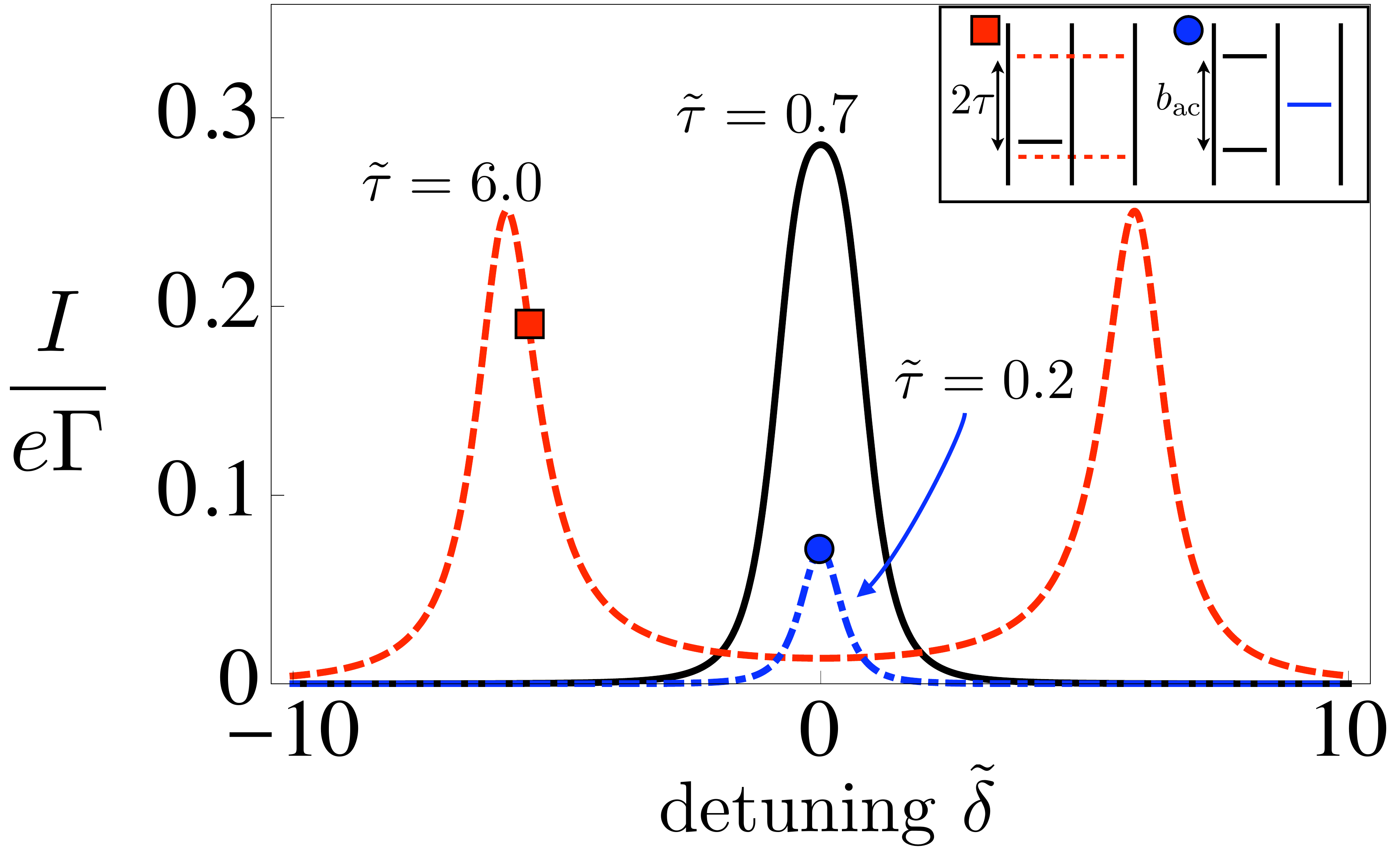}
\caption{\label{fig:results}
(Color online)
Current as function of the dimensionless detuning $\tilde \delta = \delta / b_{\rm ac}$
of the driving frequency from resonance, for
different values of the dimensionless tunnel amplitude 
$\tilde \tau = \tau / b_{\rm ac}$.
Inset: Energy diagrams corresponding to the two indicated points of the main plot.
}
\end{figure}

Using Eq.~\eqref{eq:resonantcurrent}, in Fig.~\ref{fig:results} we plot the current as a 
function of dimensionless detuning $\tilde \delta = \delta/ b_{\rm ac}$
for three different values of the dimensionless tunneling amplitude
$\tilde \tau = \tau / b_{\rm ac}$. 
The three curves correspond to three different regimes:
(i) For $\tilde \tau = 6.0$ the current shows two Lorentzian peaks as the function of 
detuning, which have width $b_{\rm ac}/\sqrt{2}$ and maximal current $e\Gamma/4$
in the entire $\tilde \tau \gg 1$ regime.
In this situation, hybridization of $\mid \!\! K'_L,0 \rangle$ and 
$\mid \!\! 0,K_R \rangle$ results in a `bonding' and an `antibonding' state having
energies $\mp \tau$, depicted as dashed lines in the left diagram of the inset of 
Fig. \ref{fig:results}.
The ac field can be resonant with only one of these
states (in the inset it is the bonding one),
as the current peak width is set by the Rabi energy $b_{\rm ac}$,
which is significantly smaller than the peak separation $2\tau$ in this case.
This parameter regime seems to be suitable for EVR detection as the 
resonant current is
large, i.e., comparable to the tunneling rate of the lead-dot barriers
(see, however, estimate of the background current $I_0$ below). 
(ii) For $\tilde \tau = 0.7$, the energy separation $2\tau$ between the two peaks of 
regime (i) becomes comparable to the Rabi energy $b_{\rm ac}$, and therefore
the two peaks merge into a single one.
This parameter setting is optimal for EVR detection as the peak 
value of the current is $2e\Gamma/7$, exceeding that in regime (i). 
(iii) For $\tilde \tau = 0.2$, the Rabi energy $b_{\rm ac}$ becomes dominant over
interdot tunneling $\tau$. 
This implies that the minimal energy distance between the two (1,0) states 
cannot drop below $b_{\rm ac}$, and 
current is maximal if they
hybridize equally ($\propto \tau/b_{\rm ac}$) with $\mid \!\! 0,K_R \rangle$
(this maximal-current situation is shown in the right diagram of the inset
of Fig. \ref{fig:results}).
This $\tilde\tau \ll 1$ regime is less suitable for EVR detection as the peak 
current is only a small fraction $2\tilde\tau^2 \Gamma$ of the
barrier transparency $\Gamma$.
The line shape is approximately a Lorentzian for small 
detuning, with width $b_{\rm ac}/\sqrt{5}$.

To take a numerical example, we use again
$b_{\rm ac} = 1.3\, \mu$eV,
$\Dso = 370\, \mu$eV, $b_L = b_R = 32.5\,\mu$eV, $\phi_R-\phi_L = \pi/3$,
which yields $\tau \approx 0.075 t$.
This implies that by tuning $t$ to $90.2\, \mu$eV, $10.5\, \mu$eV and $3.0\, \mu$eV,
one arrives to the previously plotted and described situations (i), (ii) and (iii), 
respectively.
Furthermore, it is reasonable to assume a lead-dot tunneling rate 
$\Gamma = 62.5$ MHz, which translates to a measurable current scale 
$e\Gamma \approx 10$ pA,
and a tunneling-induced level broadening of $\hbar \Gamma\approx 40$ neV 
$\ll b_{\rm ac},\tau$ in all cases.
We also estimate the EVR detection visibility by comparing the background current 
$I_0$ [see Eq. \eqref{eq:i0}], 
i.e., the current in the absence of the ac electric field, to the
current at resonant EVR driving. 
In the cases (i), (ii) and (iii), the background current
 $I_0$  is estimated from Eq. \eqref{eq:i0}
as 1.2 pA, 16 fA and 1 fA, respectively. 
For this numerical example, we can conclude that in cases (ii) and (iii) the 
background current is much smaller than the EVR contribution due to the ac 
electric field ($\sim 1$ pA from Fig. \ref{fig:results} and $e\Gamma = 10$ pA), 
and therefore EVR could indeed be detected in this transport measurement.

Valley relaxation, e.g., mediated by valley mixing and accompanied 
by phonon emission, can affect the dynamics of the EVR. 
By including valley relaxation in our transport model, 
we have found that the results shown in Fig. \ref{fig:results} do not change 
significantly as long as the valley relaxation rate is below
$b_{\rm ac}/\hbar$ and $\tau/\hbar$.
In principle, EVR could also be detected by generalizing
the proposed setup to pulsed-gated transport or charge sensing 
experiments
\cite{Koppens-esr,Nowack-esr,Laird-prl-edsr,Churchill-13cntprl},
which may allow for measuring the valley decoherence time scale.
Finally, we note that the ideas described in this paper in the context of CNTs
could be generalized to QDs in other multivalley materials, such as silicon.

We acknowledge DFG for financial support within Grants No. SFB 767, SPP 1285, 
and FOR 912.


\begin{thebibliography}{21}
\expandafter\ifx\csname natexlab\endcsname\relax\def\natexlab#1{#1}\fi
\expandafter\ifx\csname bibnamefont\endcsname\relax
  \def\bibnamefont#1{#1}\fi
\expandafter\ifx\csname bibfnamefont\endcsname\relax
  \def\bibfnamefont#1{#1}\fi
\expandafter\ifx\csname citenamefont\endcsname\relax
  \def\citenamefont#1{#1}\fi
\expandafter\ifx\csname url\endcsname\relax
  \def\url#1{\texttt{#1}}\fi
\expandafter\ifx\csname urlprefix\endcsname\relax\def\urlprefix{URL }\fi
\providecommand{\bibinfo}[2]{#2}
\providecommand{\eprint}[2][]{\url{#2}}

\bibitem[{\citenamefont{Saito et~al.}(1998)\citenamefont{Saito, Dresselhaus,
  and Dresselhaus}}]{Saito-cntbook}
\bibinfo{author}{\bibfnamefont{R.}~\bibnamefont{Saito}},
  \bibinfo{author}{\bibfnamefont{G.}~\bibnamefont{Dresselhaus}},
  \bibnamefont{and}
  \bibinfo{author}{\bibfnamefont{M.}~\bibnamefont{Dresselhaus}},
  \emph{\bibinfo{title}{Physical Properties of Carbon Nanotubes}}
  (\bibinfo{publisher}{Imperial College Press}, \bibinfo{year}{1998}).

\bibitem[{\citenamefont{Rycerz et~al.}(2007)\citenamefont{Rycerz, Tworzydlo,
  and Beenakker}}]{Rycerz-valleytronics}
\bibinfo{author}{\bibfnamefont{A.}~\bibnamefont{Rycerz}},
  \bibinfo{author}{\bibfnamefont{J.}~\bibnamefont{Tworzydlo}},
  \bibnamefont{and} \bibinfo{author}{\bibfnamefont{C.~W.~J.}
  \bibnamefont{Beenakker}}, \bibinfo{journal}{Nat. Phys.}
  \textbf{\bibinfo{volume}{3}}, \bibinfo{pages}{172} (\bibinfo{year}{2007}).

\bibitem[{\citenamefont{Recher et~al.}(2007)\citenamefont{Recher, Trauzettel,
  Rycerz, Blanter, Beenakker, and Morpurgo}}]{Recher-graphenering}
\bibinfo{author}{\bibfnamefont{P.}~\bibnamefont{Recher}},
  \bibinfo{author}{\bibfnamefont{B.}~\bibnamefont{Trauzettel}},
  \bibinfo{author}{\bibfnamefont{A.}~\bibnamefont{Rycerz}},
  \bibinfo{author}{\bibfnamefont{Y.~M.} \bibnamefont{Blanter}},
  \bibinfo{author}{\bibfnamefont{C.~W.~J.} \bibnamefont{Beenakker}},
  \bibnamefont{and} \bibinfo{author}{\bibfnamefont{A.~F.}
  \bibnamefont{Morpurgo}}, \bibinfo{journal}{Phys. Rev. B}
  \textbf{\bibinfo{volume}{76}}, \bibinfo{pages}{235404}
  (\bibinfo{year}{2007}).

\bibitem[{\citenamefont{Kuemmeth et~al.}(2008)\citenamefont{Kuemmeth, Ilani,
  Ralph, and McEuen}}]{Kuemmeth-spinorbit-in-cnts}
\bibinfo{author}{\bibfnamefont{F.}~\bibnamefont{Kuemmeth}},
  \bibinfo{author}{\bibfnamefont{S.}~\bibnamefont{Ilani}},
  \bibinfo{author}{\bibfnamefont{D.~C.} \bibnamefont{Ralph}}, \bibnamefont{and}
  \bibinfo{author}{\bibfnamefont{P.~L.} \bibnamefont{McEuen}},
  \bibinfo{journal}{Nature} \textbf{\bibinfo{volume}{452}},
  \bibinfo{pages}{448} (\bibinfo{year}{2008}).

\bibitem[{\citenamefont{Churchill
  et~al.}(2009{\natexlab{a}})\citenamefont{Churchill, Kuemmeth, Harlow,
  Bestwick, Rashba, Flensberg, Stwertka, Taychatanapat, Watson, and
  Marcus}}]{Churchill-13cntprl}
\bibinfo{author}{\bibfnamefont{H.~O.~H.} \bibnamefont{Churchill}},
  \bibinfo{author}{\bibfnamefont{F.}~\bibnamefont{Kuemmeth}},
  \bibinfo{author}{\bibfnamefont{J.~W.} \bibnamefont{Harlow}},
  \bibinfo{author}{\bibfnamefont{A.~J.} \bibnamefont{Bestwick}},
  \bibinfo{author}{\bibfnamefont{E.~I.} \bibnamefont{Rashba}},
  \bibinfo{author}{\bibfnamefont{K.}~\bibnamefont{Flensberg}},
  \bibinfo{author}{\bibfnamefont{C.~H.} \bibnamefont{Stwertka}},
  \bibinfo{author}{\bibfnamefont{T.}~\bibnamefont{Taychatanapat}},
  \bibinfo{author}{\bibfnamefont{S.~K.} \bibnamefont{Watson}},
  \bibnamefont{and} \bibinfo{author}{\bibfnamefont{C.~M.}
  \bibnamefont{Marcus}}, \bibinfo{journal}{Phys. Rev. Lett.}
  \textbf{\bibinfo{volume}{102}}, \bibinfo{pages}{166802}
  (\bibinfo{year}{2009}{\natexlab{a}}).

\bibitem[{\citenamefont{Steele et~al.}(2009)\citenamefont{Steele, Gotz, and
  Kouwenhoven}}]{Steele-cntdqd}
\bibinfo{author}{\bibfnamefont{G.~A.} \bibnamefont{Steele}},
  \bibinfo{author}{\bibfnamefont{G.}~\bibnamefont{Gotz}}, \bibnamefont{and}
  \bibinfo{author}{\bibfnamefont{L.~P.} \bibnamefont{Kouwenhoven}},
  \bibinfo{journal}{Nat. Nanotechnol.} \textbf{\bibinfo{volume}{4}},
  \bibinfo{pages}{363} (\bibinfo{year}{2009}).

\bibitem[{\citenamefont{Bulaev et~al.}(2008)\citenamefont{Bulaev, Trauzettel,
  and Loss}}]{Bulaev-socincntdots}
\bibinfo{author}{\bibfnamefont{D.~V.} \bibnamefont{Bulaev}},
  \bibinfo{author}{\bibfnamefont{B.}~\bibnamefont{Trauzettel}},
  \bibnamefont{and} \bibinfo{author}{\bibfnamefont{D.}~\bibnamefont{Loss}},
  \bibinfo{journal}{Phys. Rev. B} \textbf{\bibinfo{volume}{77}},
  \bibinfo{pages}{235301} (\bibinfo{year}{2008}).

\bibitem[{\citenamefont{Flensberg and Marcus}(2010)}]{Flensberg-bentnanotubes}
\bibinfo{author}{\bibfnamefont{K.}~\bibnamefont{Flensberg}} \bibnamefont{and}
  \bibinfo{author}{\bibfnamefont{C.~M.} \bibnamefont{Marcus}},
  \bibinfo{journal}{Phys. Rev. B} \textbf{\bibinfo{volume}{81}},
  \bibinfo{pages}{195418} (\bibinfo{year}{2010}).

\bibitem[{\citenamefont{Nowack et~al.}(2007)\citenamefont{Nowack, Koppens,
  Nazarov, and Vandersypen}}]{Nowack-esr}
\bibinfo{author}{\bibfnamefont{K.~C.} \bibnamefont{Nowack}},
  \bibinfo{author}{\bibfnamefont{F.~H.~L.} \bibnamefont{Koppens}},
  \bibinfo{author}{\bibfnamefont{Y.~V.} \bibnamefont{Nazarov}},
  \bibnamefont{and} \bibinfo{author}{\bibfnamefont{L.~M.~K.}
  \bibnamefont{Vandersypen}}, \bibinfo{journal}{Science}
  \textbf{\bibinfo{volume}{318}}, \bibinfo{pages}{1430} (\bibinfo{year}{2007}).

\bibitem[{\citenamefont{Laird et~al.}(2007)\citenamefont{Laird, Barthel,
  Rashba, Marcus, Hanson, and Gossard}}]{Laird-prl-edsr}
\bibinfo{author}{\bibfnamefont{E.~A.} \bibnamefont{Laird}},
  \bibinfo{author}{\bibfnamefont{C.}~\bibnamefont{Barthel}},
  \bibinfo{author}{\bibfnamefont{E.~I.} \bibnamefont{Rashba}},
  \bibinfo{author}{\bibfnamefont{C.~M.} \bibnamefont{Marcus}},
  \bibinfo{author}{\bibfnamefont{M.~P.} \bibnamefont{Hanson}},
  \bibnamefont{and} \bibinfo{author}{\bibfnamefont{A.~C.}
  \bibnamefont{Gossard}}, \bibinfo{journal}{Phys. Rev. Lett.}
  \textbf{\bibinfo{volume}{99}}, \bibinfo{pages}{246601}
  (\bibinfo{year}{2007}).

\bibitem[{\citenamefont{Ando and Nakanishi}(1998)}]{ando-aob1}
\bibinfo{author}{\bibfnamefont{T.}~\bibnamefont{Ando}} \bibnamefont{and}
  \bibinfo{author}{\bibfnamefont{T.}~\bibnamefont{Nakanishi}},
  \bibinfo{journal}{J. Phys. Soc. Jpn.} \textbf{\bibinfo{volume}{67}},
  \bibinfo{pages}{1704} (\bibinfo{year}{1998}).

\bibitem[{\citenamefont{P\'alyi and Burkard}(2010)}]{Palyi-cnt-spinblockade}
\bibinfo{author}{\bibfnamefont{A.}~\bibnamefont{P\'alyi}} \bibnamefont{and}
  \bibinfo{author}{\bibfnamefont{G.}~\bibnamefont{Burkard}},
  \bibinfo{journal}{Phys. Rev. B} \textbf{\bibinfo{volume}{82}},
  \bibinfo{pages}{155424} (\bibinfo{year}{2010}).

\bibitem[{\citenamefont{Ando}(2000)}]{Ando-spinorbit}
\bibinfo{author}{\bibfnamefont{T.}~\bibnamefont{Ando}}, \bibinfo{journal}{J.
  Phys. Soc. Jpn.} \textbf{\bibinfo{volume}{69}}, \bibinfo{pages}{1757}
  (\bibinfo{year}{2000}).

\bibitem[{\citenamefont{Jeong and Lee}(2009)}]{Jeong-soi-cnt}
\bibinfo{author}{\bibfnamefont{J.-S.} \bibnamefont{Jeong}} \bibnamefont{and}
  \bibinfo{author}{\bibfnamefont{H.-W.} \bibnamefont{Lee}},
  \bibinfo{journal}{Phys. Rev. B} \textbf{\bibinfo{volume}{80}},
  \bibinfo{pages}{075409} (\bibinfo{year}{2009}).

\bibitem[{\citenamefont{Izumida et~al.}(2009)\citenamefont{Izumida, Sato, and
  Saito}}]{Izumida-soiincnt}
\bibinfo{author}{\bibfnamefont{W.}~\bibnamefont{Izumida}},
  \bibinfo{author}{\bibfnamefont{K.}~\bibnamefont{Sato}}, \bibnamefont{and}
  \bibinfo{author}{\bibfnamefont{R.}~\bibnamefont{Saito}}, \bibinfo{journal}{J.
  Phys. Soc. Jpn.} \textbf{\bibinfo{volume}{78}}, \bibinfo{pages}{074707}
  (\bibinfo{year}{2009}).

\bibitem[{\citenamefont{Jhang et~al.}(2010)\citenamefont{Jhang, Marganska,
  Skourski, Preusche, Witkamp, Grifoni, van~der Zant, Wosnitza, and
  Strunk}}]{Jhang-cntspinorbit}
\bibinfo{author}{\bibfnamefont{S.~H.} \bibnamefont{Jhang}},
  \bibinfo{author}{\bibfnamefont{M.}~\bibnamefont{Marganska}},
  \bibinfo{author}{\bibfnamefont{Y.}~\bibnamefont{Skourski}},
  \bibinfo{author}{\bibfnamefont{D.}~\bibnamefont{Preusche}},
  \bibinfo{author}{\bibfnamefont{B.}~\bibnamefont{Witkamp}},
  \bibinfo{author}{\bibfnamefont{M.}~\bibnamefont{Grifoni}},
  \bibinfo{author}{\bibfnamefont{H.}~\bibnamefont{van~der Zant}},
  \bibinfo{author}{\bibfnamefont{J.}~\bibnamefont{Wosnitza}}, \bibnamefont{and}
  \bibinfo{author}{\bibfnamefont{C.}~\bibnamefont{Strunk}},
  \bibinfo{journal}{Phys. Rev. B} \textbf{\bibinfo{volume}{82}},
  \bibinfo{pages}{041404} (\bibinfo{year}{2010}).

\bibitem[{\citenamefont{Churchill
  et~al.}(2009{\natexlab{b}})\citenamefont{Churchill, Bestwick, Harlow,
  Kuemmeth, Marcos, Stwertka, Watson, and Marcus}}]{Churchill-cntspinblockade}
\bibinfo{author}{\bibfnamefont{H.~O.~H.} \bibnamefont{Churchill}},
  \bibinfo{author}{\bibfnamefont{A.~J.} \bibnamefont{Bestwick}},
  \bibinfo{author}{\bibfnamefont{J.~W.} \bibnamefont{Harlow}},
  \bibinfo{author}{\bibfnamefont{F.}~\bibnamefont{Kuemmeth}},
  \bibinfo{author}{\bibfnamefont{D.}~\bibnamefont{Marcos}},
  \bibinfo{author}{\bibfnamefont{C.~H.} \bibnamefont{Stwertka}},
  \bibinfo{author}{\bibfnamefont{S.~K.} \bibnamefont{Watson}},
  \bibnamefont{and} \bibinfo{author}{\bibfnamefont{C.~M.}
  \bibnamefont{Marcus}}, \bibinfo{journal}{Nat. Phys.}
  \textbf{\bibinfo{volume}{5}}, \bibinfo{pages}{321}
  (\bibinfo{year}{2009}{\natexlab{b}}).

\bibitem[{\citenamefont{P\'{a}lyi and Burkard}(2009)}]{Palyi-spinblockade}
\bibinfo{author}{\bibfnamefont{A.}~\bibnamefont{P\'{a}lyi}} \bibnamefont{and}
  \bibinfo{author}{\bibfnamefont{G.}~\bibnamefont{Burkard}},
  \bibinfo{journal}{Phys. Rev. B} \textbf{\bibinfo{volume}{80}},
  \bibinfo{pages}{201404(R)} (\bibinfo{year}{2009}).


\bibitem[{\citenamefont{Mason et~al.}(2004)\citenamefont{Mason, Biercuk, and
  Marcus}}]{Mason-cntdqds}
\bibinfo{author}{\bibfnamefont{N.}~\bibnamefont{Mason}},
  \bibinfo{author}{\bibfnamefont{M.~J.} \bibnamefont{Biercuk}},
  \bibnamefont{and} \bibinfo{author}{\bibfnamefont{C.~M.}
  \bibnamefont{Marcus}}, \bibinfo{journal}{Science}
  \textbf{\bibinfo{volume}{303}}, \bibinfo{pages}{655} (\bibinfo{year}{2004}).

\bibitem[{\citenamefont{Koppens et~al.}(2006)\citenamefont{Koppens, Buizert,
  Tielrooij, Vink, Nowack, Meunier, Kouwenhoven, and
  Vandersypen}}]{Koppens-esr}
\bibinfo{author}{\bibfnamefont{F.~H.~L.} \bibnamefont{Koppens}},
  \bibinfo{author}{\bibfnamefont{C.}~\bibnamefont{Buizert}},
  \bibinfo{author}{\bibfnamefont{K.~J.} \bibnamefont{Tielrooij}},
  \bibinfo{author}{\bibfnamefont{I.~T.} \bibnamefont{Vink}},
  \bibinfo{author}{\bibfnamefont{K.~C.} \bibnamefont{Nowack}},
  \bibinfo{author}{\bibfnamefont{T.}~\bibnamefont{Meunier}},
  \bibinfo{author}{\bibfnamefont{L.~P.} \bibnamefont{Kouwenhoven}},
  \bibnamefont{and} \bibinfo{author}{\bibfnamefont{L.~M.~K.}
  \bibnamefont{Vandersypen}}, \bibinfo{journal}{Nature}
  \textbf{\bibinfo{volume}{442}}, \bibinfo{pages}{766} (\bibinfo{year}{2006}).

\end{thebibliography}
\end{document}